# Time-correlated electron and photon counting microscopy

Sotatsu Yanagimoto[1], Naoki Yamamoto[1], Tatsuro Yuge[2], Hikaru Saito[1, 3], Keiichirou Akiba[1, 4, *], Takumi Sannomiya [1, †]

[1] Department of Materials Science and Engineering, School of Materials and Chemical Technology, Tokyo Institute of Technology, 4259 Nagatsuta, Midoriku, Yokohama, 226-8503, Japan

[2] Department of Physics, Shizuoka University, Shizuoka, Shizuoka, 422-8529, Japan

[3] Department of Integrated Materials, Institute for Materials Chemistry and Engineering, Kyushu University, 6-1 Kasugakoen, Kasuga, Fukuoka 816-8580, Japan

[4] Takasaki Institute for Advanced Quantum Science, National Institutes for Quantum Science and Technology, 1233 Watanuki, Takasaki, Gunma, 370-1292, Japan

---

* akiba.keiichiro@qst.go.jp

† sannomiya.t.aa@m.titech.ac.jp

# Abstract

Electron microscopy based on high-energy electrons allows nanoscopic analytical imaging taking advantage of secondarily generated particles. Especially for cathodoluminescence, the correlation between primary incident electrons and emitted photons includes information on the entire interaction process. However, electron-photon time correlation tracking the relaxation dynamics of luminescent materials has so far not been achieved. In this work, we propose time-correlated electron and photon counting microscopy, where coincidence events of primary electrons and generated photons are counted after interaction. The electron-photon time correlation enables extracting a unique lifetime of the emitter independent of the photon state, accounting for coherent and incoherent photon generation processes. We also introduce a correlation factor and discuss the correlation between electrons and generated coherent photons. Through momentum selection, we observe correlation changes indicating the presence of pair correlation originated from the electron-photon entanglement. The present work lays the foundation for developing next-generation electron microscopy based on quantum correlation.

# Introduction

Electron microscopy has been utilized for nanoscopic or even down to atomic imaging of various materials such as semiconductors, ceramics, polymers, and biological tissues or molecules. The spatial resolution of the state-of-the-art transmission electron microscopes has already reached scales well below the Bohr radius thanks to the short de Broglie wavelength of the accelerated electrons[1]. Besides the extreme spatial resolution, nanoscopic analysis by electron microscopy extends to visualizing material functions including optical/photonic properties[2–4]. For instance, portions of the primary electrons are inelastically scattered by the sample and give information on the electronic state of the constituent materials through electron energy loss spectroscopy (EELS)[5]. The loss energy is used for the generation of secondary (quasi-)particles such as electrons, plasmons or photons. These secondary particles are also used to analyze the sample, such as in energy dispersive x-ray spectroscopy (EDS)[6], Auger electron spectroscopy (AES)[7], and cathodoluminescence (CL) spectroscopy[8,9]. These secondarily generated particles are correlated to the primary electrons through various parameters in the interaction process, *e.g.* time, energy or momentum.

Among these various analysis methods, time correlation measurement of the secondarily generated photons (CL) has recently been introduced using a quantum optical method of Hanbury Brown-Twiss (HBT) interferometry[10]. The obtained time correlation between CL photons was successfully used for the deep-subwavelength characterization of the nonclassical feature of light[10,11]. Unless a single quantum emitter is excited by electron-beam, the photon correlation of CL typically shows a large photon-bunching[12], a phenomenon in which photons in the light beam are preferably close to each other, which is not observed in photoluminescence for the same type of photon sources. This bunching effect can be utilized for measuring the lifetime of the nanostructured emitters[13–15], providing additional functions to electron microscopy. The correlation not only between the generated photons but also between the interacted primary electrons and the generated photons (or secondary particles) can provide or select information of the interaction process. For instance, the correlations of electrons and generated X-rays[16–18] or photons[19] have proven to improve the signal-to-noise ratio of collected data. Varkentina *et al.* demonstrated evaluation of carrier relaxation pathways using coincidence measurements of the inelastically scattered electrons of selected energies triggered by photon emission events[20]. However, they do not extensively discuss the relaxation dynamics, and the temporal resolution of their system is insufficient for evaluating most luminescence materials, such as semiconductors while abundant information on the interaction processes is expected to be available in the dynamics.

In this study, we demonstrate time-correlated electron and photon counting microscopy (TEPCoM), where the coincidence of the incident electron transmitted through the sample and the generated photon is counted using time correlation measurement based on the scanning transmission electron microscopy (STEM) CL. This is a direct (albeit stochastic) coincidence detection of a photon induced by an accelerated electron and the electron that produced this photon. This method accomplishes an analysis approach for photon generation processes widely applicable to any emitters, not requiring photon-bunching from the sample nor active pulsing of the electron beam, which greatly facilitates the experimental setup compared to time-resolved CL instruments with a pulsed electron gun. With a parabolic mirror of the CL setup and by incorporating detectors with the time resolution of sub-nanoseconds, this setup is applicable to the evaluation of various luminescent materials. Additionally, by demonstrating electron-photon pair detection with momentum selection, we investigated an enhancement of a parameter associated with quantum correlation, which is expected to provide technological concepts including entanglement generated through inelastic interaction with materials or cavities[21,22]. From this perspective, the proposed TEPCoM approach is one of the candidates for next-generation electron microscopy that utilizes quantum correlation[23] and quantum information experiments using free electrons[24].

## Results and Discussion

### Experimental setup

For the TEPCoM setup, we modified a CL-HBT interferometry system used for photon-photon correlation function $g_{\text{pp}}^{(2)}(\tau)$ measurement by replacing one of the single-photon detectors with a transmitted electron detector to obtain the electron-photon correlation function $g_{\text{ep}}^{(2)}(\tau)$, as schematically illustrated in Fig. 1(a, b). In the electron detector, transmitted electrons are converted to photons by a scintillator so that a time-resolved photon counter can be applied, as shown in Fig. 1(c). To evaluate the electron-photon correlation, time correlation histograms as a function of the time delay $\tau$ between the photon and electron detector signals are acquired after collecting sufficient counts for statistics (Fig. 1(d-f)). We normalize the correlation signals by the number of counts at $\tau \to \infty$, similarly to the second order photon correlation function commonly used in the HBT setup[14]. The correlation values at the zero delay $g_{ep}^{(2)}(0)$ above one indicates the detection of correlated electron-photon pairs. This electron-photon coincidence means that the electron that generated the photon detected at the photon detector is specifically captured at the electron detector. While conventional HBT measurement provides an auto-correlation function, where the photons emitted from the same source are detected, TEPCoM provides a cross-correlation function for different sources corresponding to the electron-photon correlations. We also realized simplified electron-photon correlation measurement using a sample-integrated electron detector instead of the dedicated electron detector, which can be readily used even in scanning electron microscopy.

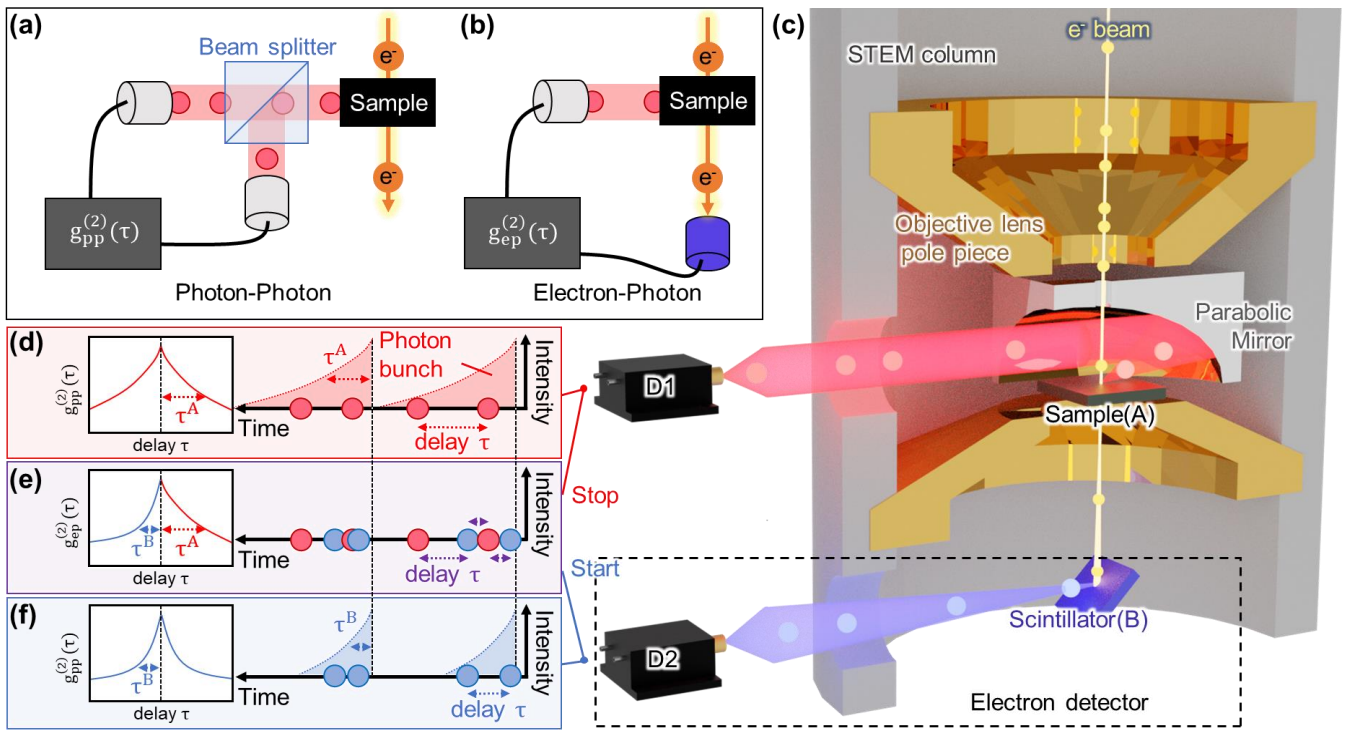

**Fig. 1 Illustration of time-correlated electron and photon counting microscopy (TEPCoM).**

(a, b) Schematic diagrams of the CL-HBT system for the photon-photon correlation measurement (a) and the proposed electron-photon correlation measurement (b). The photon-photon correlation function $g_{\text{pp}}^{(2)}(\tau)$ is obtained by measuring the time correlation between the generated photons using a beam splitter while the electron-photon correlation function $g_{\text{ep}}^{(2)}(\tau)$ provides the correlation between the photons from the sample and the excitation electrons with an electron detector. (c) Illustration of TEPCoM measurement system. Emitted photons from the sample (material A) are collected through a parabolic mirror and counted by a single-photon counting module D1. For electron detection, the scintillator (material B) converts the electrons to photons, which are counted by another single-photon counting module D2. The signals from D1 and D2 are introduced to the start and stop channels of the correlator to obtain the time correlation function. (d-f) Schematics of the photon detection time trace from the sample and scintillator, where the shaded triangles correspond to the photon bunch. A symmetric correlation curve is obtained for the auto-correlation of each detector (d, f) as a function of time delay $\tau$, while an asymmetric curve is obtained for the correlation between the photon (D1) and electron (D2) detectors (e), corresponding to the lifetime of the sample and scintillator.

**Formulation of correlation function**

To derive the formalism of the electron-photon correlation, we consider the cross-correlation function of two luminescent materials A and B, i.e. sample and scintillator, assuming multiple two-level systems. The detected CL photon intensity (count per time) from A (B) at time $t$ is denoted by $I^{A(B)}(t)$. When the signal from B is set as the trigger (start), the normalized cross-correlation function $g^{(2)}_{AB}(\tau)$ can be defined as

$$g^{(2)}_{AB}(\tau) = \frac{\langle I^A(t+\tau) I^B(t) \rangle}{\langle I^A(t+\tau) \rangle \langle I^B(t) \rangle}, \tag{1}$$

where $\langle \cdots \rangle$ means ensemble average[25]. By using the radiation efficiency $p_r^{A(B)}$ and detection efficiency of the detector $\alpha^{A(B)}$ for A (B), $I^{A(B)}(t)$ is expressed as

$$I^{A(B)}(t) = p_r^{A(B)} \alpha^{A(B)} n^{A(B)}(t), \tag{2}$$

where $n^{A(B)}(t)$ is the number of the excited emitters. Since the constant coefficients are canceled in the $g^{(2)}_{AB}(\tau)$ calculation, the derivation of the function $g^{(2)}_{AB}(\tau)$ comes down to the calculation of $n^{A(B)}(t)$. Let us consider the i-th electron incidence event at time $t_i$, where one incident electron generates $N_i^{A(B)}$ excited emitters at each excitation. Between time $t_i$ and $t_{i+1}$, no electron arrives, and thus the rate equation from a single excitation event of material A (B) can be described as

$$\frac{dn_i^{A(B)}(t)}{dt} = -\frac{1}{\tau^{A(B)}} n_i^{A(B)}(t), \tag{3}$$

where $\tau^{A(B)}$ is the lifetime of the material A (B). The solution including $N_e$ times of the electron incidence event over sufficiently long time is given by a sum of each decay process as

$$n^{A(B)}(t) = \sum_{i=1}^{N_e} N_i^{A(B)} e^{-\frac{(t-t_i)}{\tau^{A(B)}}} \theta(t - t_i), \tag{4}$$

where $\theta(t - t_i)$ is a Heaviside step function: $\theta(t - t_i) = 0$ for $t < t_i$, $\theta(t - t_i) = 1$ for $t \geq t_i$. Note that some of $N_i^{A(B)}$ are zero within the $N_e$ events. We now consider two luminescent materials A and B as the sample and the scintillator, as shown in Fig. 1(c), respectively. Since an electron is efficiently converted to photons by the scintillator (material B), counting the converted photon generation event corresponds to counting the event of the transmitted electron arrival. Therefore, we rewrite the subscripts for materials A and B as photon (p) and electron (e), respectively, and the electron-photon correlation function in this study is derived as

$$g_{ep}^{(2)}(\tau) = 1 + \frac{e\xi^{ep}}{(\tau^e + \tau^p)I_b}\left[e^{-|\tau|/\tau^e}\theta(-\tau) + e^{-|\tau|/\tau^p}\theta(\tau)\right] \quad (5)$$

using the electron beam current $I_b$ and elemental charge $e$ (see the derivation details in Supplementary Note 1). At $\tau = 0$, we define $\theta(-0) = 0$ and $\theta(+0) = 1$. We introduced an excitation correlation factor $\xi^{ep}$ describing the correlation between the number of excitations of the sample (p) and scintillator (e):

$$\xi^{ep} = \lim_{N_e \to \infty} \frac{\frac{1}{N_e}\sum_{i=1}^{N_e} N_i^e N_i^p}{\left(\frac{1}{N_e}\sum_{i=1}^{N_e} N_i^e\right)\left(\frac{1}{N_e}\sum_{i=1}^{N_e} N_i^p\right)} = \frac{\langle N^e N^p \rangle}{\langle N^e \rangle \langle N^p \rangle}. \quad (6)$$

Equation (5) shows that the electron-photon correlation function is asymmetric. The decay functions in the positive ($\tau > 0$) and negative ($\tau < 0$) time delay range directly give the lifetime of the sample and scintillator, respectively. Since the decay functions of the two materials are separately obtained for $\tau > 0$ and $\tau < 0$, the scintillator lifetime does not affect the lifetime measurement of the sample of interest.

Since the detected photons and the electron that generated the photons are always time-correlated, $g_{ep}^{(2)}(0) > 1$ is observed for any photon state emitted from the sample, indicating no limitation in the sample, unlike previous CL-HBT requiring clear photon (anti-)bunching[10,12]. In addition, this correlation measurement by TEPCoM provides the excitation correlation factor $\xi^{\mathrm{ep}}$, formulating the strength of the correlation. If the sample and scintillator are independently excited, $\xi^{\mathrm{ep}}$ should be unity. However, we keep this factor as a variable since the experimentally obtained values may vary and can be used to evaluate the correlation, which is discussed later.

## Electron-photon correlation in incoherent CL

We first demonstrate the $g^{(2)}_{\mathrm{ep}}(\tau)$ measurement of the incoherent CL from 100 nm nanodiamond particles (NDPs) with 900 nitrogen-vacancy (NV) centers per particle, which exhibit photon-bunching[12,14], as illustrated in Fig. 2(a). The lifetime of these NDP samples obtained from $g^{(2)}_{\mathrm{pp}}(\tau)$ is $\tau^{\mathrm{NDP}} \sim 20$ ns[14]. We performed the measurements with two scintillators with different lifetimes as the electron-photon converters: bulk Y$_2$SiO$_5$:Ce (YSO) and polyvinyl toluene-based plastic scintillator. The lifetime values are $\tau^{\mathrm{YSO}} = 50.5 \pm 0.1$ ns for YSO and $\tau^{\mathrm{plastic}} = 2.89 \pm 0.08$ ns for the plastic scintillator (see Supplementary Note 2 for details). Figure 2 shows the electron-photon correlation results with asymmetric time correlation curves with respect to $\tau = 0$, unlike symmetric bunching curves of the conventional HBT. The exponential fittings of the positive and negative time regimes, as shown in Fig. 2(b), gave the decay time of $\tau^{\mathrm{neg}} = 25.0 \pm 3.7$ ns and $\tau^{\mathrm{pos}} = 54.0 \pm 6.2$ ns respectively, where the signal from the NDP sample is sent to the start channel and the YSO scintillator signal to the stop. The decay time values in the negative and positive regime correspond well to the lifetime values of the photon source (NDP) and the scintillator (YSO). The asymmetric correlation curve is flipped by switching the signals introduced into the two channels as shown in Fig. 2(c). The decay time values also flip to $\tau^{neg} = 49.9 \pm 5.5$ ns and $\tau^{pos} = 17.7 \pm 2.8$ ns. Such asymmetric and switchable behaviors with separated decays in the negative and positive sides are in good agreement with the prediction of Eq. (5), proving the successful electron-photon correlation measurement. When the plastic scintillator with a shorter lifetime is used, the decay time of the electron detector side is drastically reduced, giving $\tau^{neg} = 20.1 \pm 2.2$ ns and $\tau^{pos} = 2.18 \pm 0.56$ ns for the start as the NDP (photon detector) and stop as the plastic scintillator (electron detector), as shown in Fig. 2(d), and $\tau^{neg} = 3.51 \pm 0.70$ ns and $\tau^{pos} = 22.0 \pm 2.2$ ns when the channels are switched (Fig. 2(e)). The decay time value well agrees with the lifetime of the material of each channel. These results with two scintillators show that the emission lifetime of the scintillator does not affect the shape of the decay curve of the sample side as expected from Eq. (5), ensuring the applicability of TEPCoM to the lifetime measurement. This electron-photon correlation approach is advantageous over the conventional auto-correlation HBT setup when applied to weak luminescent materials because photon counts are not lost by the beam splitter and electron counting, which is the event triggering the photon generation, does not degrade the performance of photon counting.

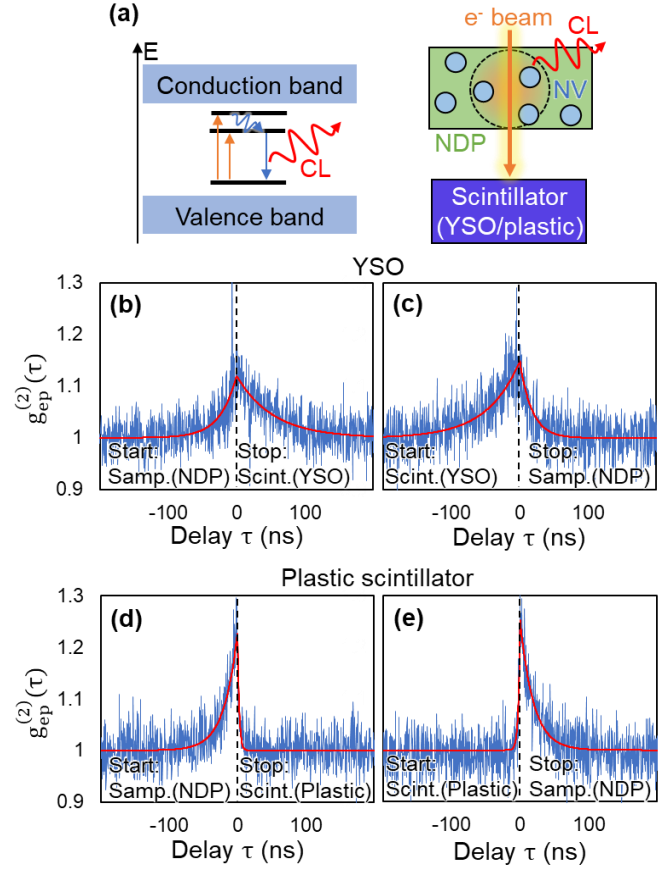

**Fig. 2 Electron-photon correlation measurement in incoherent CL using nanodiamond particles (NDPs) with two different scintillators.** (a) Schematic illustration of incoherent luminescence of NV centers in diamond. The excited NV centers relax through phonon emission into defect levels before light emission. (b, c) and (d, e) show the electron-photon correlation measurement of NDPs using YSO (long lifetime) and plastic (short lifetime) scintillators, respectively. The exponential decay curves fitted by Eq. (5) are overlaid on the measured results (blue) as solid red lines. Correlation curves in (b) and (d) are acquired with the photon detector signal introduced to the start channel and the electron detector signal to the stop channel. For (c) and (e), the start and stop channels are switched.

Here, we also evaluate the excitation correlation factor $\xi^{ep}$, which can be calculated without any fitting by integrating the experimentally obtained $g^{(2)}_{ep}(\tau)$ curve through the following equation,

$$\int_{-\infty}^{\infty} (g^{(2)}_{ep}(\tau) - 1)\, d\tau = \frac{e\xi^{ep}}{I_b}. \tag{7}$$

$\xi^{ep}$ can be obtained independently from other fitting-based parameters such as lifetime. The mean values obtained in Fig. 2 are $\xi^{ep} = 0.225 \pm 0.005$ for the YSO scintillator and $\xi^{ep} = 0.170 \pm 0.004$ for the plastic scintillator. These values are smaller than one because the detected signals are averaged over large scan areas which include places without photon emission. When the detection signal includes uncorrelated background signals, the spatially averaged $\xi^{ep}$ value falls below one (see Supplementary Note 3 and 4 for the actual scan area and background noise effect, respectively).

To assess the feasibility of nanoscale lifetime mapping as well as the spatial dependence of $\xi^{ep}$, we performed electron-photon correlation mapping, as shown in Fig. 3. We used a NDP sample and a YSO scintillator for the mapping. The STEM bright-field image and panchromatic CL images of the NDPs sample are shown in Fig. 3 (a), (b) and (c). We set the signal from the NDP sample to the start channel and the YSO scintillator signal to the stop. $g^{(2)}_{ep}(0) > 1$, meaning the detection of excitation electron and generated photon pairs, is observed in most areas in the $g^{(2)}_{ep}(0)$ plot of Fig. 3(d). The error (Fig. 3(e)) is smaller where the CL intensity (Fig. 3(c)) is higher. The decay time of the negative and positive regimes agree with the lifetime of NDP and YSO, respectively, as shown in Fig. 3(f, g). The lifetime of the YSO scintillator, which should not depend on the electron-beam position, is homogeneously distributed with around 50 ns values except for the places with large errors due to low photon signals. The result shown in Fig. 3(f, g) confirms the applicability of the electron-photon correlation mapping to visualize the lifetime distribution at the nanoscale. The excitation correlation factor $\xi^{ep}$ is close to unity, where the NDP photon emission gives sufficient signals compared to the dark count of the detectors or the stray light. Indeed, in the area without photon emission, the $\xi^{ep}$ value becomes almost zero due to dominant noise signals. This confirms that $\xi^{ep}$ values in the results of Fig.2 are degraded due to signal-averaging over empty spaces. When properly detected, $\xi^{ep}$ becomes unity for such incoherent CL without specific correlation.

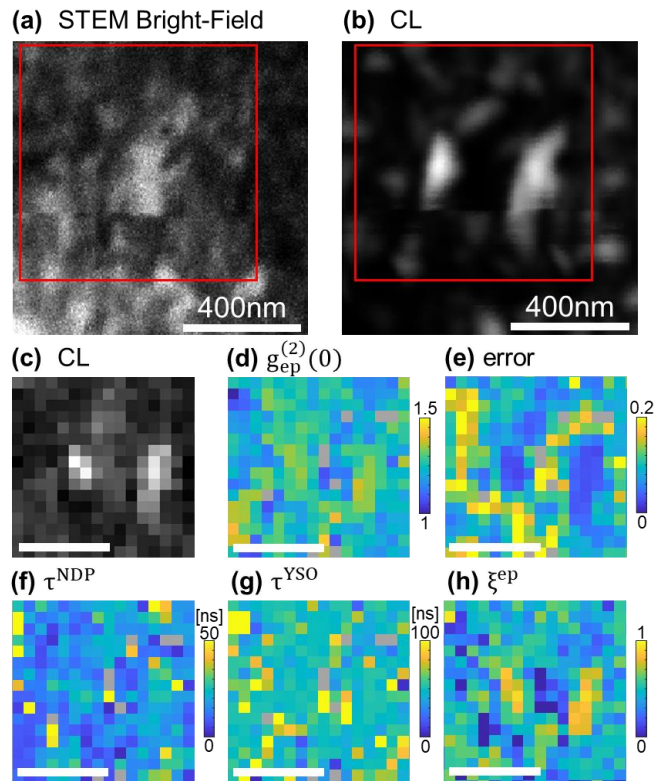

**Fig. 3 Electron-photon correlation mapping.** Nanodiamond particles (NDPs) and YSO are the sample and scintillator, respectively. The photon signal from the NDPs is sent to the start channel

and the YSO scintillator signal to the stop. (a) STEM bright-field image, (b) panchromatic cathodoluminescence (CL) image of NDPs sample, and (c-h) the distributions obtained from the correlation map measurement and integrated 40s per pixel. The mapping area is indicated as the red square in (a) and (b). (c) A panchromatic CL image of the squared area, (d) a map of $g_{ep}^{(2)}(0)$ and (e) fitting error represented by the 95% confidence interval of the $g_{ep}^{(2)}(0)$. (f-g) show the decay times of the negative and positive regimes corresponding to the lifetimes of NDP and YSO, respectively. (h) The excitation correlation function $\xi^{ep}$. The scale bars in panels c-h correspond to 400 nm.

## Electron-photon correlation in coherent CL

For the measurement of the coherent emission, we adopt gold nanoparticles (AuNP) with a diameter of about 200 nm, as shown in the SEM and the panchromatic images in Fig. 4(b) and (c). We confirmed that the emission of AuNP is dominantly from the coherent surface plasmons with negligible contribution from the inter-band transition (see Supplementary Note 5 for details). While incoherent CL, as shown above for NDP with NV centers or semiconductor[12,26], loses direct information of the excitation electrons through the relaxation process, coherent CL can conserve the phase relationship between excitation and radiation[4], momentum of the excitation[27], and total energy of electrons and photons[28], as illustrated in Fig. 4(a). Therefore, the correlation function obtained in this section reflects the coincidence detection of electron-photon pairs which conserve coherence. The coherent CL typically shows a significantly short lifetime compared to incoherent one. AuNPs also show fast decay of its surface plasmon resonance typically in the range of $10^{-14}$ sec[29], which is extremely fast and cannot be evaluated with the time resolution of the current instrument. However, we are still able to evaluate the correlation factor $\xi^{ep}$ from the correlation histogram even for such fast emissions (see Supplementary Note 6 for details). For this measurement, the YSO scintillator is used as the electron detector. Figure 4(d, e) shows the correlation curves, giving an asymmetric shape with a sharp drop/rise only on the sample side. When the start channel is set to the photon detector (AuNP sample) and the stop channel to the YSO electron detector (Fig. 4(d)), the exponential fitting gave the decay times of $\tau^{neg} = 0.92 \pm 0.92$ ns and $\tau^{pos} = 63.1 \pm 10.6$ in the negative (AuNP) and positive (scintillator) time ranges, respectively. The $\tau$-positive side's decay time agrees with the lifetime of YSO. For the negative side, the measured decay time is reaching the instrument limit (see Supplementary Note 7 for the instrument response function). However, Eq. (5) suggests that the decay for $\tau < 0$ reasonably reflects the surface plasmon emission lifetime. By switching the signals introduced into the start and stop channels, as shown in Fig. 4(e), the correlation curve is flipped, resulting in decay time values $\tau^{neg} = 45.6 \pm 6.9$ ns and $\tau^{pos} = 2.19 \pm 1.10$ ns. Finally, the spatially averaged $\xi^{ep}$ value for this coherent emission was $0.478 \pm 0.011$, lower than unity, which is because the signal from the vacuum regime is included in the scan similarly to the previous discussion for the incoherent CL.

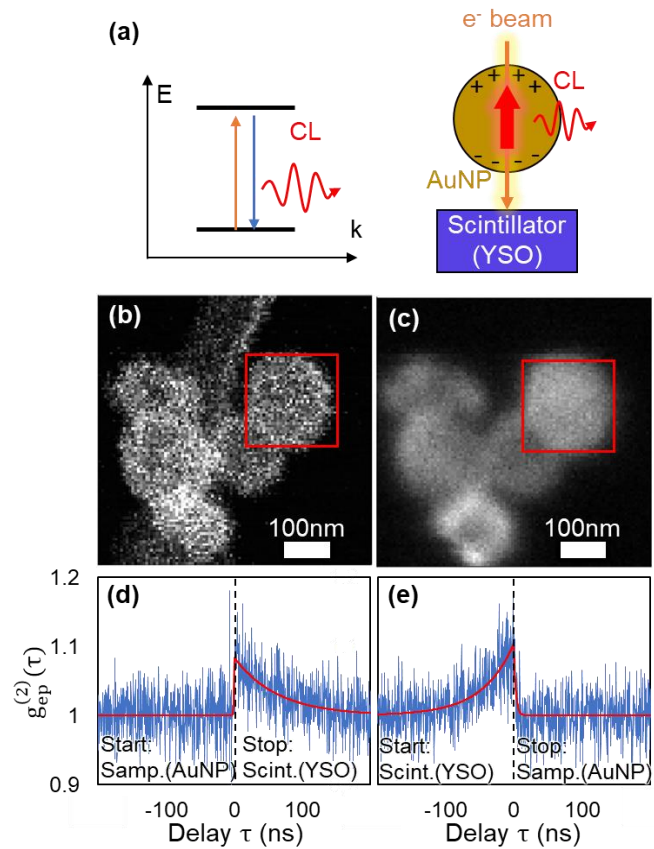

**Fig. 4 Electron-photon correlation measurement for coherent CL using AuNPs.** (a) Schematic illustration of coherent cathodoluminescence (CL) from a AuNP. Fast electrons with the electromagnetic near field of optical frequencies coherently excites an electric dipole in the AuNP. The emitted photon and excitation electron conserve the momentum, energy, and phase. (b) SEM image, (c) panchromatic image of AuNPs. The red squares indicate the electron beam scan areas for the correlation measurement. (d) Correlation curve with the coherent photon signal and the electron detector signal from the YSO scintillator respectively introduced into the start and stop channels. (e) Correlation curve with the switched channel configuration from (d). The fitted curves (red) are overlaid on the raw data (blue).

The incident electron and generated photon in the coherent process are expected to be entangled through energy and momentum conservations. Since the parameter $\xi^{ep}$ describes the correlation related to such electron and photon pairs, we can consider that Eq. (6) is also applicable to quantum description. Here, we discuss the pair correlation associated with entanglement between electrons and photons. Assuming that only the electron-photon scattering events are selectively detected (e.g., by momentum selection) and that an electron is converted to an $N$-photon state by the scintillator for simplicity, we can write the detected state as

$$|\psi\rangle = \frac{1}{\sqrt{1+\eta}}(|0\rangle_e|0\rangle_p + \sqrt{\eta}|N\rangle_e|1\rangle_p), \quad (0 < \eta \ll 1), \tag{8}$$

where $\eta$ is the excitation probability of coherent CL and $|n\rangle_{e(p)}$ is the $n$-number state of the detected scintillator (sample) photon mode. From Eq. (8), $\xi^{ep} = 1 + 1/\eta > 1$ can be deduced, which is consistent with the interpretation that coincident excitation events of the sample and the scintillator favorably occur rather than independent excitation events. In contrast, the state without the selection, where all the electrons are detected, can be described as

$$|\psi'\rangle = \frac{1}{\sqrt{1+\eta'}}(|N\rangle_e|0\rangle_p + \sqrt{\eta'}|N\rangle_e|1\rangle_p) = \frac{1}{\sqrt{1+\eta'}}|N\rangle_e(|0\rangle_p + \sqrt{\eta'}|1\rangle_p), \tag{9}$$

which leads to $\xi^{ep} = 1$. Thus, the selective detection of the correlated electron-photon pair, such as momentum selection, is expected to change the correlation factor $\xi^{ep}$.

Here, we experimentally perform momentum-selected electron-photon correlation measurements to see if the selection gives influence on $\xi^{ep}$. A plasmonic nanohole array with a hole diameter of 250 nm and a hole center-center distance of 500 nm is adopted as the sample for this measurement, as shown in Fig. 5(a). The electron diffraction patterns of the sample are shown in Fig. 5(b) and (c). The momentum selection in the photon detection is achieved by inserting a mask in the light path, which limits the emission angle within half of the upper hemisphere[30], as schematically illustrated in Fig. 5(d) and (e). The electron momentum selection is performed by limiting the scattering angle using an aperture edge on the diffraction plane (see Methods and Supplementary Note 8 for the details). The aperture is adjusted so that the detected electron counts are the same for the two electron detection conditions. The electron diffraction pattern reflecting the nanohole array structure (Fig. 5(c)) indicates that the electron momentum can be resolved in the scale of photon momentum in the visible; the scattering angle of the recoiling electron by horizontal photon emission of a wavelength of 500 nm is about 5.7 μrad. We note that the states decohere before the detection and cannot directly show the entanglement with this sample and the electron detector.

Applying Eq. (7) to the obtained correlation curves (Fig. 5(d) and (e)), the excitation correlation factor for the momentum-conserved pair is obtained as $\xi^{ep} = 0.211 \pm 0.004$ and that for the non-conserved pair as $\xi^{ep} = 0.172 \pm 0.004$. The value of $\xi^{ep}$ is enhanced by the momentum selection, which indicates that the electron-photon correlated pair can be more effectively detected. The smaller absolute value of $\xi^{ep}$, compared to the previous result in Fig. 4, is because only the low-angle-scattered electrons are used in the momentum selection measurement, discarding the high-angle scattered electrons by small features, such as atomic potentials, which count for more than 70% of the incident electrons. This significantly lowers the effective signal compared to the uncorrelated backgrounds. These discarded electrons have also excited (or possibly even more preferentially excited) photons, resulting in a decrease of absolute $\xi^{ep}$ values, as well as causing the $g^{(2)}_{ep}(0) > 1$ feature even in the non-momentum-conserved detection. Portions of electrons or photons that have transferred the momentum to the sample also become the background signals reducing $\xi^{ep}$, although these are equally present in both detection conditions due to the sample symmetry. Nevertheless, the demonstrated change of $\xi^{ep}$ through momentum selection ensures an essential contribution from the momentum-correlated electrons and photon pairs.

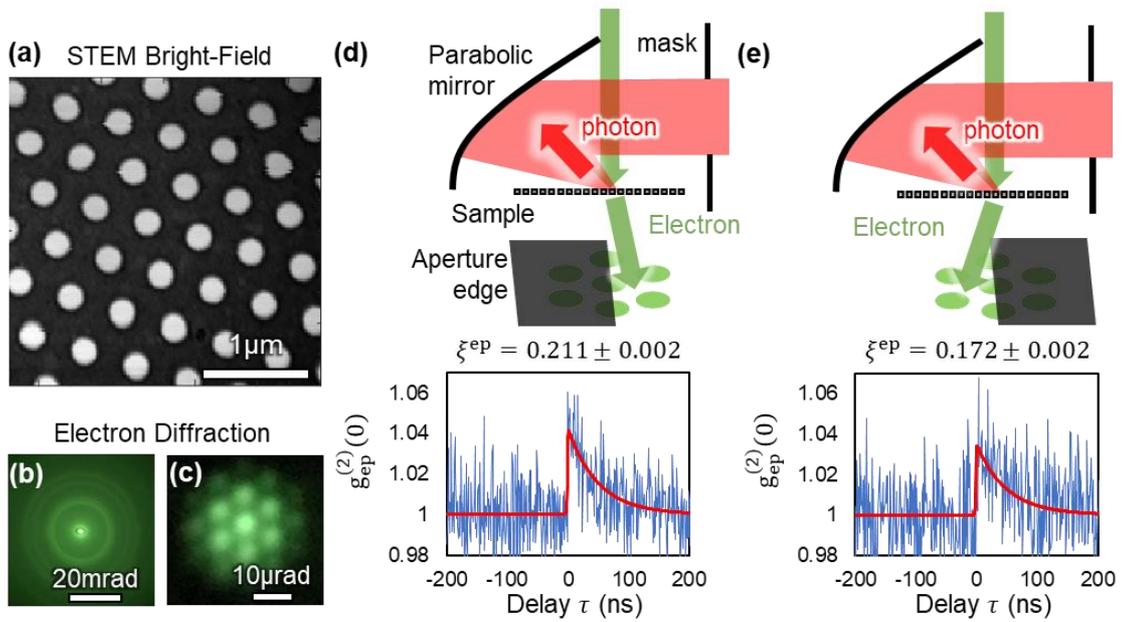

**Fig. 5 Momentum-resolved electron-photon correlation measurement with a silver nanohole array.** (a) STEM bright-field image. The pitch interval of the holes is 500 nm and the hole diameter is 250 nm. The array film consists of a 50 nm silver film deposited on a suspended 200 nm SiN membrane. Panels (b) and (c) show the electron diffraction patterns with conventional and ultra-long camera lengths, respectively. The momentum selection is performed in the long

camera length condition with the illumination beam diameter of 7 μm. (d,e) Schematics and correlation curves of the (d) momentum-conserved and (e) non-conserved) measurements performed by selecting half of the diffracted beams using an aperture edge for electron detection and by detecting the photons emitted within the corresponding (or opposite) half of the upper hemisphere angles. The fitted curves (red) are superposed on the raw data (blue). The momentum-conserved correlation (d) gives an enhanced correlation factor of $\xi^{ep} = 0.211 \pm 0.004$ compared to the non-conserved one (e) of $\xi^{ep} = 0.172 \pm 0.004$. A 700 nm short-pass optical filter is inserted to avoid slight inclusion of the filament emission in the photon detection. The aperture edge is adjusted so that the transmitted electron beam current is reduced by 70% for both electron detection conditions.

**Electron-photon correlation using built-in scintillator substrate**

We have shown the electron-photon correlation, where the sample and electron detectors are spatially separated approximately by 1 m. This requires a special STEM setup for the electron detector (Fig. 1(c)). As a simpler alternative, which can easily be performed in SEM, we here demonstrate a system with the electron detector scintillator integrated in the sample substrate, which allows electron-photon correlation measurement by a conventional HBT-CL setup without a dedicated electron detector. This is realized by using optical wavelength filters to energy-separate the photons of the sample from those from the scintillator. A CrPbBr$_3$ (CPB) substrate is used as the electron detector, and the NDPs as the measurement target are dropped on the CPB substrate. The illustration of the setup is shown in the Fig. 6(a), and the emissions from these two materials are separable by a 550 nm short-wavelength pass (SP) filter for CPB and a 561 nm long-wavelength pass (LP) filter for NDP, as shown in the emission spectra in the Fig. 6(b). The lifetime of semiconductor CPB is $\tau^{CPB} \sim 1 ns$[31], fast enough for a scintillator (see Supplementary Note 9 for details). Figure 6(c) shows the measured correlation function when the emissions from CPB is set to the start channel and NDP to the stop. This result with an asymmetric correlation curve demonstrates the electron-photon correlation, in the same manner as the separated system (Figs. 2-5). The positive and negative decay functions are $\tau^{neg} = 1.73 \pm 0.73$ ns and $\tau^{pos} = 14.1 \pm 2.4$ ns, well corresponding to the lifetime of CPB and NDP, respectively. These features indicate that the events of NDP and CPB emissions excited by the same electron are selectively captured. Thus, we have shown that the electron-photon correlation measurement can also be realized with a simpler setup based on the conventional HBT-CL system.

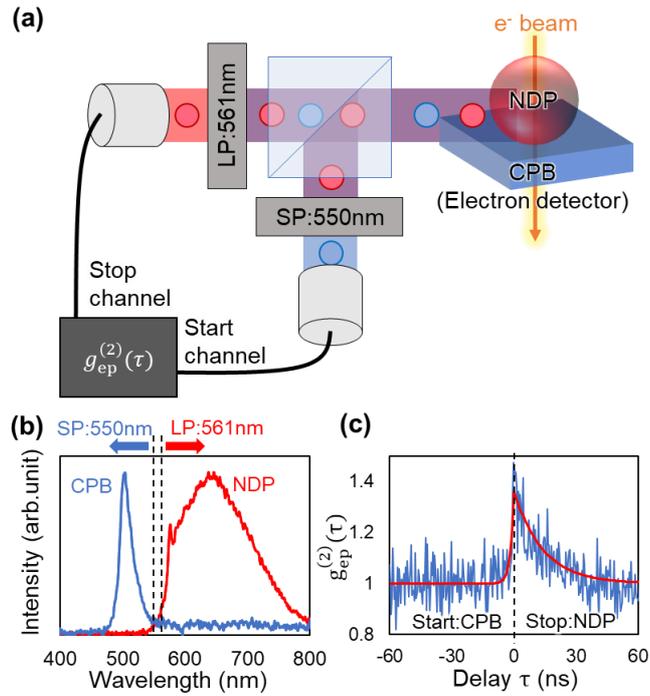

**Fig. 6 Electron-photon correlation measurement using a built-in electron-detector substrate.**

(a) Schematics of the measurement setup based on the photon-photon correlation CL-HBT setup. Photons from the sample (NDP) and the electron detector (CPB) are separated by the photon wavelength using short pass (SP) and long pass (LP) filters. (b) The normalized spectra of CPB (blue) and NDP (red) show that the 550 nm-SP and 561nm-LP effectively separate the two emission signals. (c) Correlation curve measured with the start channel connected to the electron detector (CPB) and the stop to the sample (NDP). The fitted curve (red) is overlaid on the raw data (blue).

## Conclusions

We have proposed and successfully demonstrated time-correlated electron and photon counting microscopy, where single coincidence events of the excitation electron and generated photons are statistically evaluated. The absence of a beam splitter in our setup provides an advantage over conventional HBT setups when applied to weak luminescent materials. In principle, we can capture all electrons passing through the sample and record the time traces of all the photons induced by the electrons. The electron-photon correlation curve exhibits asymmetric decay curves with the correlation value at the zero-time delay exceeding unity for both coherent and incoherent CL. The derived analytical formulation of the correlation function predicts that the CL lifetime of the sample is obtained independently from the lifetime of the scintillator of the electron detector. We confirmed this analytical prediction experimentally by using two scintillator materials with different lifetimes. Thus, our study provides a theoretical framework for coincidence detection measurements of electron-photon pairs and proposes its application for evaluating the emission lifetime without pulsing the electron beam. This approach is applicable to any luminescent material regardless of the emitted photon state. By synchronizing the detection with the electron beam scan, visualization of the emission lifetime distribution is also possible at the nanoscale far below the diffraction limit of light. By integrating a scintillator substrate to the sample, the electron-photon correlation measurement can be performed without a dedicated electron detection unit. The analytical formulation revealed that the electron-photon correlation provides the excitation correlation factor $\xi^{\mathrm{ep}}$, which is enhanced by momentum selection in coherent CL, demonstrating an effective contribution of the momentum-correlated electron-photon pairs. Such experimentally obtained correlation of the interacted electron and photon pair for the coherent CL is a concrete step forward to the detection of their quantum entanglement, where energy, momentum, and phase relations are conserved between the electrons and photons. For this perspective, energy selection of the electron would further reinforce the approach. The presented scheme of the correlation measurement sets a road map for the future microscopy techniques selectively imaging the interaction processes[20,32] or quantum light sources[33,34] as well as for potential quantum communication design using electrons[35].

## Methods

**Details of the Instruments.** The experimental TEPCoM setup is based on a scanning transmission electron microscope (STEM) (JEM-2000FX, JEOL, Japan) with a thermionic-emission gun modified for CL detection, as schematically illustrated in Fig. 1(c). In the photon detection part, a parabolic mirror collimates the CL emission from the sample and transfers the photons to the single photon counting module (SPCM) (PDM, MPD, Italy). For the electron detection part, we modify the conventional STEM detection system so that the electron detection signals from the scintillator, i.e. photons, are forwarded to the second SPCM. The time correlation statistics is acquired by the correlation electronics (TimeHarp260 PICO, Pico Quant, Germany) receiving the electric pulse signals from the two SPCMs. All measurements are performed at room temperature with an acceleration voltage of 80 kV for the experiments of Figs. 2-4,6, and 160kV for Fig. 5. The error shown in this work corresponds to 95 % confidence (two-sigma) interval.

For the momentum- (angle-) resolved electron detection, we used a dedicated electron optics setup to achieve a long camera length (more details are in the Supplementary Note 8). The illumination area is set to ~ 7 μm in diameter and the illumination half angle ~ 1.5 μrad (Fig. 5c). The detection half angle for the momentum-resolved measurement is ~ 20 μrad, and the half of this detection disk is covered by the aperture edge to select the momentum of a certain direction.

## Data availability

All data needed to evaluate the conclusions are present in the main text and the Supplementary Information. Other data that support the findings of this study are available from the corresponding authors upon reasonable request.

## Acknowledgements

This work was financially supported by JST PRESTO (JPMJPR17P8), JSPS KAKENHI (21K18195, 22H01963, 22H05032, 23KJ0892), JST FOREST (JPMJFR213J) and Research Foundation for Opto-Science and Technology. The authors are grateful to Dr. K. Nakamura for providing us experimental equipment.

## Author contributions

TS, HS and KA conceived the main concept of the electron-photon correlation. SY formulated the electron-photon correlation function with the help of NY, TY, KA and TS. SY, NY, KA and TS developed the measurement setup. SY, NY, KA and TS performed experiments. HS contributed to the sample preparation. SY, NY, KA and TS wrote the manuscript. All authors discussed the results and contributed to the manuscript.

## Corresponding authors

Keiichirou Akiba* :　akiba.keiichiro@qst.go.jp

Takumi Sannomiya† :　sannomiya.t.aa@m.titech.ac.jp

## Supplementary information

Supplementary Notes 1-9.

## Competing interests

The authors declare no competing interests.

# Supplementary Information

# Supplementary Note 1:

# DERIVATION OF ASYMMETRIC CROSS-CORRELATION FUNCTION $g^{(2)}_{AB}(\tau)$

To derive the asymmetric correlation function, we consider that the target material A contains enough luminescent centers. As previously demonstrated and calculated, when a single incident electron excites multiple luminescent centers of the material simultaneously, a photon bunching phenomenon occurs[1]. The 2$^{nd}$ order autocorrelation function of the emitted photons from material A, $g^{(2)}_{A(auto)}(\tau)$, obtained from the conventional CL-HBT experiment, has a symmetric functional form[1,2]. Let us now consider an event in which a single incident electron simultaneously excites two materials A and B, both of which contain a large number of emission centers. In this case, the 2$^{nd}$ order cross-correlation function $g^{(2)}_{AB}(\tau)$, which is the time correlation between the photons emitted from A and B, shows "bunching" and gives asymmetric electron-photon correlation functions. Now we assume a situation where the electron beam current is weak enough that an excitation process of one incident electron has no influence on the next electron excitation.

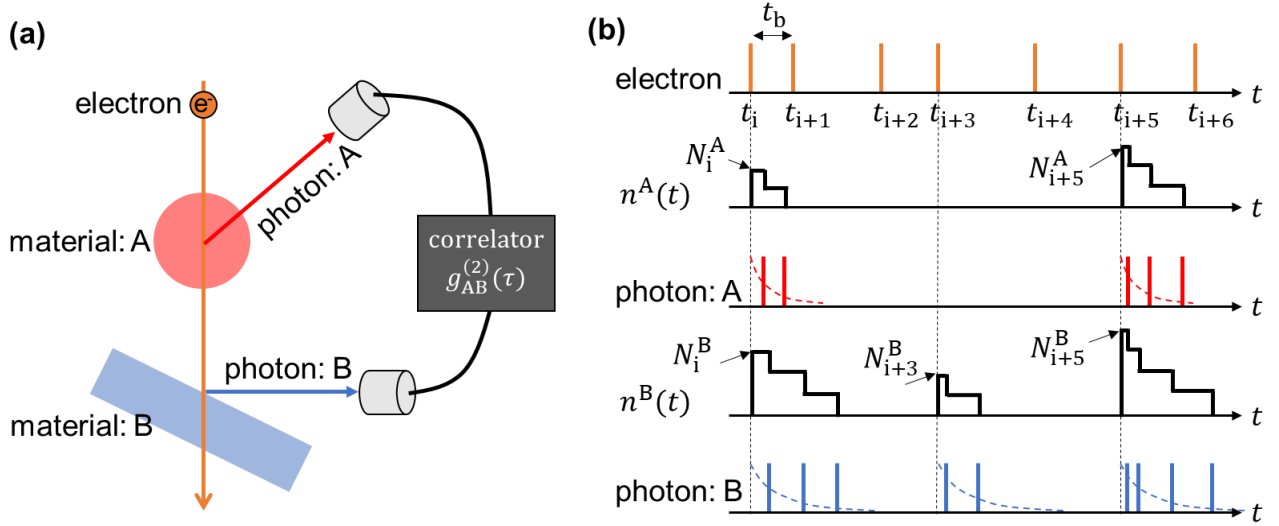

**Fig. S1 Concept of the electron-photon correlation.** (a) Schematic diagram of electron-photon correlation measurement. A single fast electron excites materials A and B almost simultaneously, and their time correlations are measured. (b) Classical picture of the relation among electron incidence, number of excited emission centers $n^{A(B)}(t)$ and photon emission from the materials A and B.

For the electron beam current $I_b$, the average time interval $t_b$ between consecutive electron incidences is given by $t_b = (I_b/e)^{-1}$. Using the number of events $N_e$ of fast electrons incident onto the sample during the measurement time $T$, the following equation holds:

$$\lim_{T \to \infty} \frac{N_e}{T} = \frac{1}{t_b}. \tag{S1}$$

First, we consider the excitation and relaxation processes of material A. Let $G^A(t)$ be a function that expresses the excitation of material A. It can be written as

$$G^A(t) = \sum_{i=1}^{N_e} N_i^A \delta(t - t_i), \tag{S2}$$

S3

where $N_i^A$ is the number of emission centers that are simultaneously excited at time $t_i$. Using the average excitation number of emission centers per electron $N_{ex}^A$, the average excitation rate $G^A$ is given by

$$G^A = \langle G^A(t) \rangle = \frac{N_{ex}^A}{t_b}. \tag{S3}$$

If the number of centers at the excited state at time $t$ is $n^A(t)$, the rate equation is expressed as

$$\frac{dn^A(t)}{dt} = G^A(t) - \gamma^A n^A(t), \tag{S4}$$

where $\gamma^A = \gamma_r^A + \gamma_{nr}^A$ is the total decay rate considering the radiative and non-radiative decay paths. Between the time $t_i$ and $t_{i+1}$, no excitation occurs, thus the rate equation from a single excitation event can be described as

$$\frac{dn^A(t)}{dt} = -\gamma^A n^A(t), \tag{S5}$$

and the solution of the single process with the initial condition that $n_i^A(0) = N_i^A$ at $t = t_i = 0$, is given by

$$n^A(t) = N_i^A e^{-\gamma^A t}. \tag{S6}$$

This situation is schematically illustrated in Fig. S1. The single excitation-emission process is repeated for each randomly incident electron. When $N_e$ electron incidence events occur over a sufficiently long-time range, the total number $n^A(t)$ is expressed by

$$n^A(t) = \sum_{i=1}^{N_e} N_i^A e^{-\gamma^A (t-t_i)} \theta(t - t_i), \tag{S7}$$

where $\theta(t - t_i)$ is a step function, i.e., $\theta(t - t_i) = 1$ for $t \geq t_i$ and $0$ for $t < t_i$. In the same manner, the number of excited centers in material B, $n^B(t)$, is described as

$$n^B(t) = \sum_{i=1}^{N_e} N_i^B e^{-\gamma^B(t-t_i)} \theta(t - t_i). \tag{S8}$$

Since the photon emission intensity from the sample A (B) is proportional to $n^{A(B)}(t)$, the 2$^{nd}$ order cross-correlation function $g_{AB}^{(2)}(\tau)$ at detection time delay $\tau$ for photons from A and B is defined as

$$g_{AB}^{(2)}(\tau) = \frac{\langle I^A(t+\tau) I^B(t) \rangle}{\langle I^A(t+\tau) \rangle \langle I^B(t) \rangle} = \frac{\langle n^A(t+\tau) n^B(t) \rangle}{\langle n^A(t+\tau) \rangle \langle n^B(t) \rangle}. \tag{S9}$$

According to the ergodic theorems, $\langle n^{A(B)}(t) \rangle$ can be calculated as the time average of $n^{A(B)}(t)$:

$$\langle n^A(t) \rangle = \lim_{T \to \infty} \frac{1}{T} \int_0^T \sum_{i=1}^{N_e} N_i^A e^{-\gamma^A(t-t_i)} \theta(t - t_i) dt = \lim_{T \to \infty} \frac{1}{T} \sum_{i=1}^{N_e} N_i^A \int_{-t_i}^{T-t_i} e^{-\gamma^A t} \theta(t) dt$$

$$= \lim_{T \to \infty} \frac{1}{T} \sum_{i=1}^{N_e} N_i^A \frac{1}{\gamma^A} = \frac{N_{ex}^A}{\gamma^A t_b}, \tag{S10}$$

$$\langle n^B(t) \rangle = \frac{N_{ex}^B}{\gamma^B t_b}. \tag{S11}$$

The inverse Fourier transform of $n^{A(B)}(t)$ is defined as

$$n^{A(B)}(t) = \frac{1}{2\pi} \int_{-\infty}^{\infty} n_\omega^{A(B)}(\omega) e^{-i\omega t} d\omega. \tag{S12}$$

Using Wiener-Khinchin theorem, the correlation function $\langle n^A(t+\tau) n^B(t) \rangle$ is calculated as follows

$$\langle n^A(t+\tau) n^B(t) \rangle = \lim_{T \to \infty} \frac{1}{T} \int_{-T}^{T} n^A(t+\tau) n^B(t) dt$$

$$= \lim_{T \to \infty} \frac{1}{T} \int_{-T}^{T} dt \frac{1}{2\pi} \int_{-\infty}^{\infty} n_\omega^B(\omega') e^{i\omega' t} d\omega' \frac{1}{2\pi} \int_{-\infty}^{\infty} n_\omega^A(\omega) e^{i\omega(t+\tau)} d\omega$$

$$= \lim_{T\to\infty} \frac{1}{T}\int_{-T}^{T} dt \, \frac{1}{2\pi}\int_{-\infty}^{\infty} n_\omega^B(-\omega')e^{-i\omega' t}d\omega' \, \frac{1}{2\pi}\int_{-\infty}^{\infty} n_\omega^A(\omega)e^{i\omega(t+\tau)}d\omega$$

$$= \lim_{T\to\infty} \frac{1}{T}\int_{-\infty}^{\infty} d\omega' \, \frac{1}{2\pi}\int_{-\infty}^{\infty} d\omega \, n_\omega^A(\omega) \, n_\omega^B(-\omega') \, e^{i\omega\tau} \left[\frac{1}{2\pi}\int_{-T}^{T} dt \, e^{i(\omega-\omega')t}\right]$$

$$= \lim_{T\to\infty} \frac{1}{2\pi}\int_{-\infty}^{\infty} \frac{1}{T} n_\omega^A(\omega) n_\omega^B(-\omega)e^{i\omega\tau} d\omega. \tag{S13}$$

Furthermore, considering $n_\omega^{A(B)}(\omega)$ the Fourier transform of $n^{A(B)}(t)$ defined as

$$n_\omega^A(\omega) = \int_{-\infty}^{\infty} n^A(t)e^{-i\omega t}dt = \frac{1}{\gamma^A + i\omega}\sum_{k=1}^{N_e} N_k^A e^{i\omega t_k}, \tag{S14}$$

$$n_\omega^B(-\omega) = \frac{1}{\gamma^B - i\omega}\sum_{k=1}^{N_e} N_k^B e^{-i\omega t_k}. \tag{S15}$$

Inserting (S14) and (S15) to (S13), and after some algebra, we obtain

$$\langle n^A(t+\tau)n^B(t)\rangle = \frac{\xi^{AB} N_{ex}^A N_{ex}^B}{(\gamma^A + \gamma^B)t_b}\left(e^{-\gamma^B|\tau|}\theta(-\tau) + e^{-\gamma^A|\tau|}\theta(\tau)\right) + \frac{N_{ex}^A N_{ex}^B}{\gamma^A \gamma^B t_b^2}. \tag{S16}$$

Finally, $g_{AB}^{(2)}(\tau)$ is derived as follows

$$g_{AB}^{(2)}(\tau) = 1 + \frac{\gamma^A \gamma^B \xi^{AB} t_b}{(\gamma^A + \gamma^B)}\left(e^{-\gamma^B|\tau|}\theta(-\tau) + e^{-\gamma^A|\tau|}\theta(\tau)\right), \tag{S17}$$

where $\xi^{AB}$ is the excitation correlation factor and defined as

$$\xi^{AB} = \lim_{N_e\to\infty} \frac{\frac{1}{N_e}\sum_{k=1}^{N_e} N_k^A N_k^B}{\left(\frac{1}{N_e}\sum_{k=1}^{N_e} N_k^A\right)\left(\frac{1}{N_e}\sum_{k=1}^{N_e} N_k^B\right)}. \tag{S18}$$

This factor theoretically equals one when the excitations of A and B are independent. Here, Eq. (S17) can be rewritten as

$$g_{AB}^{(2)}(\tau) = 1 + \frac{e\xi^{AB}}{(\tau^A + \tau^B)I_b}\left[e^{-|\tau|/\tau^B}\theta(-\tau) + e^{-|\tau|/\tau^A}\theta(\tau)\right], \tag{S19}$$

where $\tau^{A(B)}$ is the lifetime of the A(B) and defined as $\gamma^{A(B)} = 1/\tau^{A(B)}$. In the main text, A and B are substituted to e (electron) and p (photon).

It is worth noting the difference between the 2$^{nd}$ order cross-correlation function $g_{AB}^{(2)}(\tau)$ and the 2$^{nd}$ order autocorrelation function $g_{A(auto)}^{(2)}(\tau)$ in the conventional HBT experiment. Considering the two materials A and B to be identical, Eq. (S19) provides the 2$^{nd}$ order cross-correlation function of the emission from the same sample $g_{AA}^{(2)}(\tau)$. When the two materials are excited independently, $g_{AA}^{(2)}(\tau)$ does not equal to $g_{A(auto)}^{(2)}(\tau)$. The autocorrelation function $g_{A(auto)}^{(2)}(\tau)$ is described by assuming that the same number of emitters are excited by each incident electron in both materials A and B. This assumption implies that $N_k^A = N_k^B$ in Eq. (S18). To derive the 2$^{nd}$ order autocorrelation function $g_{A(auto)}^{(2)}(\tau)$, the parameter $\xi^{AB}$ should be replaced by the excitation autocorrelation factor $\xi_{auto}^A$, which is described as

$$\xi_{auto}^A = \lim_{N_e \to \infty} \frac{\frac{1}{N_e}\sum_{k=1}^{N_e}(N_k^A)^2}{\left(\frac{1}{N_e}\sum_{k=1}^{N_e}N_k^A\right)^2}. \tag{S20}$$

To see the relationship to the previously derived formula of the autocorrelation function[2], we now introduce an ensemble $\{N_l^{Ae}\}_{l=1}^{N^A}$, a subset of $\{N_k^A\}_{k=1}^{N_e}$ composed only of the actually excited events ($N_k^A \geq 1$). The number of the element of the $\{N_l^{Ae}\}_{l=1}^{N^A}$ is described as $N^A = \eta^A N_e$, where $\eta^A$ is the excitation probability. Using $N_l^{Ae}$ and $N^A = \eta^A N_e$, Eq. (S20) is rewritten as below.

$$\xi_{\text{auto}}^{\text{A}} = \lim_{N_e \to \infty} \frac{N_e}{N^{\text{A}}} \frac{\frac{1}{N^{\text{A}}}\sum_{l=1}^{N^{\text{A}}}(N_l^{\text{Ae}})^2}{\left(\frac{1}{N^{\text{A}}}\sum_{l=1}^{N^{\text{A}}} N_l^{\text{Ae}}\right)^2} = \frac{1}{\eta^{\text{A}}} \lim_{N_e \to \infty} \frac{\frac{1}{N^{\text{A}}}\sum_{l=1}^{N^{\text{A}}}(N_l^{\text{Ae}})^2}{\left(\frac{1}{N^{\text{A}}}\sum_{l=1}^{N^{\text{A}}} N_l^{\text{Ae}}\right)^2} \tag{S21}$$

This is the form of the excitation autocorrelation factor, which is proportional to $\frac{1}{\eta^{\text{A}}}$. Inserting Eq. (S21) to Eq. (S19), we obtain the autocorrelation function as

$$g_{\text{A(auto)}}^{(2)}(\tau) = \frac{e\xi_{\text{auto}}^{\text{A}}}{2\tau^{\text{A}} I_{\text{b}}} e^{-\frac{|\tau|}{\tau^{\text{A}}}} + 1, \tag{S22}$$

which is in agreement with the autocorrelation function derived in the previous study[2].

## Supplementary Note 2:

### PHOTON-PHOTON CORRELATION FUNCTION OF THE SCINTILLATORS

We used bulk $Y_2SiO_5$:Ce (YSO) and polyvinyl toluene-based plastic scintillators for the electron detectors. Here, the correlation curves of these two materials were measured by the conventional HBT measurements of CL and are shown in Fig. S2. The measurements were performed under the same conditions as described in the main text, i.e., accelerating voltage 80kV, electron beam current 4.8 pA and at room temperature. The lifetimes are $\tau^{\text{YSO}} = 50.5 \pm 0.1$ ns and $\tau^{\text{plastic}} = 2.89 \pm 0.08$ ns, and these are in good agreement with previously reported values[3] and catalog specifications.

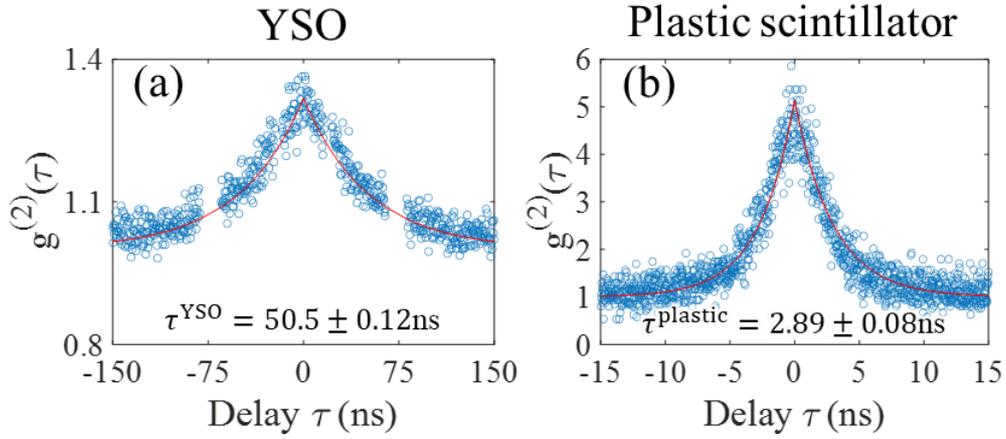

**Fig. S2 Photon-photon correlation functions for the two scintillators.** (a) and (b) show the correlation curves of YSO and plastic scintillators, respectively. In the YSO plot, the data missing around $\pm 75$ ns is caused by removing the noise due to the reflections at the end of the optical fibers, which are used for sending photons into SPCMs.

## Supplementary Note 3:

### MEASUREMENT AREA OF NANODIAMOND SAMPLES

In the measurement using the YSO and Plastic scintillators in Fig. 2 in the main text, the electron beam is scanned over the area of the nanodiamond particle (NDP) sample, as shown in the red rectangle in Fig. S3. The electron beam current is set at 4.8 pA for both measurements.

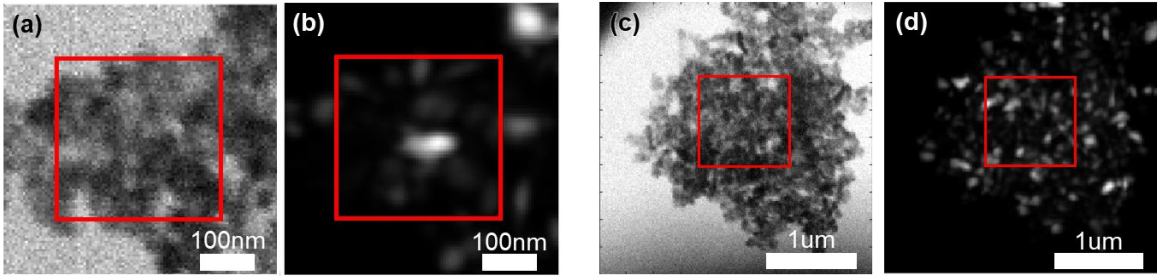

**Fig. S3 Images of the samples.** (a, b) and (c, d) are images of nanodiamond particle (NDP) samples used in measurements with YSO and Plastic scintillators, respectively. (a, c) STEM bright-field images. (b, d) Panchromatic CL images of NDP samples. The red rectangles indicate the electron beam scan areas for the correlation measurement.

## Supplementary Note 4:

### DECAY FACTOR $p_{ss}$ DUE TO BACKGROUND NOISE

In electron-photon correlation measurements, background noise such as dark counts of the detectors and ambient stray light decrease the $g^{(2)}_{AB}(0)$ value and the excitation correlation factor $\xi^{AB}$. Here, we consider the effect of such noises on the 2$^{nd}$ order cross-correlation function. The CL intensities of the material A and B with the noise are defined as

$$I^{A(B)}(t) = I_s^{A(B)}(t) + I_n^{A(B)}, \tag{S23}$$

where $I_s^{A(B)}(t)$ and $I_n^{A(B)}$ indicate the intensity of the signal and the noise, respectively. Using Eq. (S23), the cross-correlation function considering the noise effect $g^{(2)}_{AB(noise)}(\tau)$ is calculated as

$$g^{(2)}_{AB(noise)}(\tau) = \frac{\langle I^A(t+\tau)I^B(t)\rangle}{\langle I^A(t+\tau)\rangle\langle I^B(t)\rangle} = \frac{\langle I_s^A(t+\tau)I_s^B(t)\rangle + \langle I_s^A(t+\tau)I_n^B\rangle + \langle I_n^A I_s^B(t)\rangle + \langle I_n^A I_n^B\rangle}{(\langle I_s^A(t)\rangle + \langle I_n^A\rangle)(\langle I_s^B(t)\rangle + \langle I_n^B\rangle)}. \tag{S24}$$

Since the signal and noise have no temporal relation and are uncorrelated, the following equation holds:

$$\frac{\langle I_s^A(t+\tau)I_n^B\rangle}{\langle I_s^A(t+\tau)\rangle\langle I_n^B\rangle} = \frac{\langle I_n^A I_s^B(t)\rangle}{\langle I_n^A\rangle\langle I_s^B(t)\rangle} = \frac{\langle I_n^A I_n^B\rangle}{\langle I_n^A\rangle\langle I_n^B(t)\rangle} = 1. \tag{S25}$$

Using Eq. (S24) and (S24), $g^{(2)}_{AB(noise)}(\tau)$ is given by

$$g^{(2)}_{AB(noise)}(\tau) = \frac{ep_{ss}\zeta^{AB}}{(\tau^A + \tau^B)I_b}\left[e^{-|\tau|/\tau^B}\theta(-\tau) + e^{-|\tau|/\tau^A}\theta(\tau)\right] + 1. \tag{S26}$$

Here, $p_{ss}$ is the decay factor and defined as

$$p_{ss} = \frac{\langle I_s^A(t+\tau)\rangle\langle I_s^B(t)\rangle}{(\langle I_s^A(t)\rangle + \langle I_n^A\rangle)(\langle I_s^B(t)\rangle + \langle I_n^B\rangle)}. \tag{S27}$$

The decay factor $p_{ss}$ is obviously less than one, and Eq. (S26) indicates that the experimentally obtained bunching height $g^{(2)}_{AB(noise)}(0)$ decreases from the original value by $p_{ss}$. When the noise intensity $\langle I_n^{A(B)}\rangle$ is much smaller than the signal intensity $\langle I_s^{A(B)}(t)\rangle$, $p_{ss} \cong 1$ holds and Eq. (S26) equals to Eq. (S19). However, if the sample is too thick for the fast electrons to transmit, the intensity of the scintillator $\langle I_s^B(t)\rangle$ becomes as weak as the noise intensity. Alternatively, if the photon intensity from the sample is very weak, *e.g.* for too thin samples or without presence of the sample, $\langle I_s^A(t)\rangle$ may also be small compared to the noise signal. In such cases, Eq. (S27) shows that $p_{ss}$ becomes lower than one and reduces the $g^{(2)}_{AB(noise)}(0)$.

## Supplementary Note 5:

**CL SPECTRA OF THE SAMPLES**

CL spectra of nanodiamond particles (NDPs), gold nanoparticles (AuNPs) and CsPbBr$_3$ (CPB) are shown in Fig. S4. From the spectrum of the AuNP sample, one can confirm that the emission is dominantly from the coherent surface plasmons with negligible contribution from the inter-band transition .

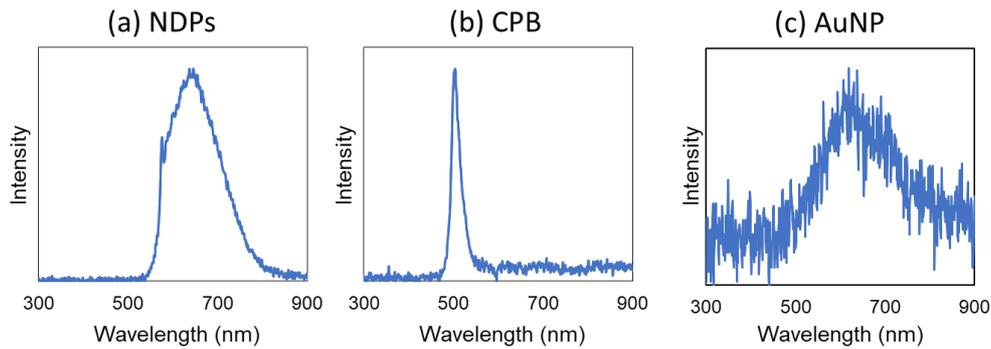

**Fig. S4 CL spectra of (a) NDPs, (b) CPB and (c) AuNP.**

## Supplementary Note 6:

**$g_{AB}^{(2)}(\tau)$ FOR COHERENT CL**

The derivation process of $g_{AB}^{(2)}(\tau)$ in Eq. (19) is based on the assumption that both materials A and B show incoherent CL. Here, we consider a sample A exhibiting coherent CL, *e.g.,* Au nanoparticle. The temporal evolution of the electric field, which is coherently excited by a fast electron, has been theoretically described in previous studies[4]. According to the temporal profile of the field, we can reasonably assume that the amplitude of the electric field rises promptly and decays exponentially. In the condition that the beam current is sufficiently small with negligible overlaps of the excitation events, the emission intensity $I^A(t)$ can be described as

$$I^A(t) \propto |E^A(t)|^2 = \sum_{k=1}^{N_e} M_k^A e^{-2\gamma^A(t-t_k)} \theta(t-t_k). \tag{S28}$$

$M_k^A$ is a dimensionless coefficient, defined as $M_k^A = 1$ when the excitation takes place and otherwise $M_k^A = 0$. The mean value of $M_k^A$ for the whole electron incident events corresponds to the excitation efficiency. Since Eq. (S7) and Eq. (S28) are identical, we can deduce an equivalent form of $\xi^{AB}$ as Eq. (S18) also for coherent CL.

## Supplementary Note 7:

**INSTRUMENT RESPONSE FUNCTION**

The correlation function is obtained as the convolution of the curves reflecting the physical properties of the sample and the instrument response function (IRF). Therefore, it is necessary to consider the IRF, especially for short-lifetime material. Our experimental setup for correlation measurement consists of two single-photon counting modules (SPCMs) (PDM, MPD, Italy) and the correlation electronics (TimeHarp260 PICO, Pico Quant, Germany). By performing HBT measurement of ultrafast pulsed laser, IRF for our system is extracted. The pulse width is 50 fs, which is four orders magnitude smaller than the specification response time of each instrument. Therefore, we can obtain IRF through this measurement.

Figure S5 shows results with the ultrafast pulse laser at two wavelengths, 400 and 800 nm. By fitting with the following equation

$$g^{(2)}(\tau) = a_1 \exp\left(-\frac{|\tau|}{\tau^1}\right) + a_2 \exp\left(-\frac{|\tau|}{\tau^2}\right) + 1, \tag{S29}$$

two overlapping exponential curves with decay times of about 52 ps and 350 ps were obtained for both wavelengths, as shown in Fig. S5. The wavelength dependence is small, and we estimate that the decay time of IRF is around 50~350 ps.

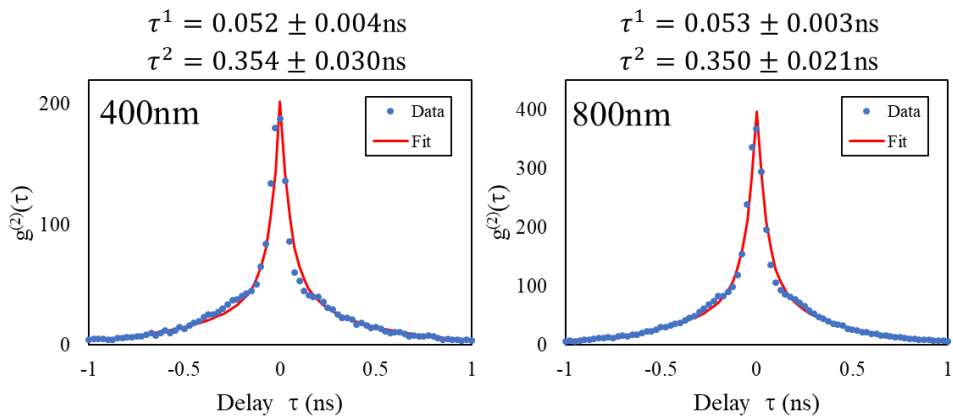

**Fig. S5 IRF of our set up.** Correlation curves measured with a pulsed laser at two wavelengths, 400 and 800 nm. The data at the two wavelengths are fitted with two overlapping exponential curves expressed by Eq. (S29).

# Supplementary Note 8:

**ELECTRON OPTICS FOR MOMENTUM RESOLVED-MEASUREMENT**

For the momentum- (angle-) resolved electron detection, we used TEM imaging mode, instead of STEM, without exciting the objective lens of the electron microscope. The ray-path is shown in Fig.S6. The illumination area (ca. 7 μm in diameter) as well as the parallelity of the incident electron beam is adjusted by the condenser lens set. The "diffraction" position (dashed line in the figure) is also controlled by the condenser lenses. This "diffraction" pattern is imaged on the selected area (SA) aperture plane by the objective mini lens placed below the sample. The magnification of the diffraction (or camera length) on the SA plane is dependent on the first diffraction position (dashed line). With this setup, momentum (angle) selection can be performed by the SA aperture. Using the magnification optics of the conventional TEM imaging optics, the final camera length of the diffraction on the screen can be set to up to a few hundred meters. The detection half angle is ~ 20 μrad, and the half of this detection disk is covered by the SA aperture edge for the momentum selection.

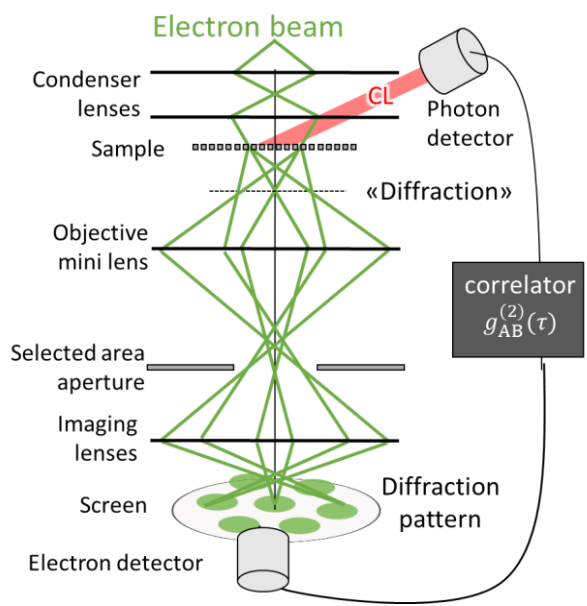

**Fig. S6. Ray path of the electron beam for the electron momentum detection.** The objective lens of the TEM is turned off and the electron optics is adjusted so that the "diffraction pattern" is imaged on the selected area (SA) aperture plane and (de)magnified on the screen by the magnification optics of the TEM imaging mode.

**Supplementary Note 9:**

**PROPERTY OF CPB FOR BUILT-IN SCINTILLATOR SYSTEM**

Integrating the "electron detector" scintillator into the sample substrate allows the electron-photon correlation measurement with a conventional HBT setup. We chose $CsPbBr_3$ (CPB) as the scintillator substrate, which is a perovskites semiconductor with high quantum efficiency[5]. The CL peak wavelength of CPB is around 530 nm and can be separated from that of NDP (600 nm) using an optical filter. We use two single-photon counting modules (SPCMs) (PerkinElmer, SPCM-AQR-14, USA) and correlation electronics (ORTEC, 9353, USA) for the correlation measurements of CPB. As shown in Fig. S7, the lifetime of the CPB is $\tau^{CPB} = 0.95 \pm 0.04$ ns, showing that this material works as a fast scintillator.

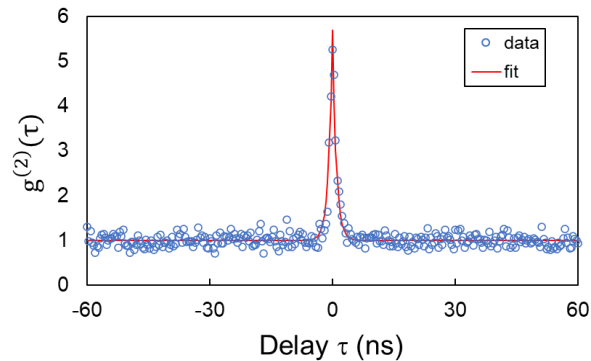

**Fig. S7 Auto-correlation curve of the CPB substrate.** The data is overlaid with the fitting curve shown by the red line. The obtained lifetime is $\tau^{CPB} = 0.95 \pm 0.04$ ns.

## Supplementary References